\documentclass[aps,prc,twocolumn,showpacs,groupedaddress,floatfix,10pt]{revtex4-1}

\usepackage{graphicx}
\usepackage{dcolumn}
\usepackage{amsfonts}

\begin{document}

\title{Microscopic Composition of Ion-Ion Interaction Potentials}
\author{A.S. Umar,$^1$ V.E. Oberacker,$^1$ J.A. Maruhn,$^2$ and P.-G. Reinhard$^3$}
\affiliation{$^1$Department of Physics and Astronomy, Vanderbilt University, Nashville, Tennessee 37235, USA \\
$^2$Institut f\"ur Theoretische Physik, Goethe-Universit\"at, D-60438 Frankfurt am Main, Germany \\
$^3$Institut f\"ur Theoretische Physik, Universit\"at Erlangen, D-91054 Erlangen, Germany}
\date{\today}


\begin{abstract}
We present a new method to calculate the total ion-ion interaction potential in terms
of building blocks which we refer to as ``single-particle interaction
potentials''.
This allows also to compose the separate contributions 
from neutrons and protons to the interaction potentials.
The method is applied to nuclear collisions via the use of time-dependent Hartree-Fock
theory. 
\end{abstract}
\maketitle

The knowledge of interaction potentials between composite many-body systems is
of fundamental importance for predicting the outcome of reactions involving
such entities.  These are often modeled using
empirical functions that depend on macroscopic variables, such as the distance
between the two centers, treating the composite objects as structureless
particles and ignoring the microscopic origins of these potentials
(e.g., Lennard-Jones potential for rare-gas atoms \cite{AM76aB},
nucleon-nucleon potentials \cite{Mac01aR}). On the
other hand, microscopic approaches try to obtain such potentials by including
the interactions of the constituents as the building-blocks and calculating
the whole potential as a function of some set of macroscopic variables
(e.g., Born-Oppenheimer approximation for molecules \cite{Wei78aB},
deformed shell models for nuclei \cite{Bra72aR}).
However, the total potential obtained this way does not reveal the
contribution coming from individual single-particle states.  Quoting
Schr\"odinger, \textit{ best knowledge of a whole does not necessarily include
  best possible knowledge of the parts}.  The entanglement of these
contributions may yield further insight as to the interaction of many-body
systems via a representative potential.

Most common microscopic approaches for calculating interaction
potentials usually employ the adiabatic or sudden approximations for
the relative motion of the interacting systems \cite{Fae87a}. Recently, we have
introduced a new microscopic approach for the calculation of ion-ion
potentials for nuclear collisions. This method is based on the
time-dependent Hartree-Fock (TDHF) description of nuclear collisions
coupled~\cite{Ne82} with a constraint on the local density at the
given instant of time.
Interaction potentials obtained using the so-called density
constrained TDHF (DC-TDHF)
method~\cite{UO06b} have been successful in describing fusion
cross-sections for a number of systems~\cite{UO09b}.  The new approach
we will describe below is generally suitable for microscopic
calculations provided Koopmans' theorem \cite{Koo33a} is applicable.

The microscopic approaches based on the mean-field approximation often use the
energy difference between the combined system and the asymptotic energies
of the individual systems to calculate the ion-ion interaction potential
as
\begin{equation}
\label{eq:vr}
 V(R)=E_{A_1+A_2}(R) - E_{A_1}^{(0)} - E_{A_1}^{(0)}\;,
\end{equation}
where $E_{A_1+A_2}(R)$ is the total energy obtained for the combined
system as a function of the distance $R$ between the two ions, and
$E_{A_i}^{(0)}$ are the individual binding energies of the two systems
calculated using the same interaction. 
The binding energies are often calculated either by integrating the energy density
over all space $ E=\int d^3 r \,{\cal H}(\mathbf{r})$ or via Koopmans' expression
$
 E=\frac{1}{2}\sum_{\alpha}w_{\alpha}(t_{\alpha}+\epsilon_{\alpha})\;,
$
where $w_{\alpha}$ denotes the occupation probability of the
single-particle state $\alpha$.  This mixture of half kinetic energies
$t_{\alpha}$ and half single-particle energies $\epsilon_{\alpha}$
applies to Hartree-Fock calculations employing purely two-body
forces.
 For mean-field calculations based on
density-dependent effective forces \cite{Ben03aR} or energy-density
functionals \cite{Dre90aB} a rearrangement term needs to be added to
the above expression \cite{RS80}. 
This can be disentangled to single-particle energies and rearrangement
term as
\begin{equation}
  E
  =
  \frac{1}{2}\sum_{\alpha}w_{\alpha}(t_{\alpha}+\epsilon_{\alpha})
  -\frac{1}{2}Tr\left(<HF|\frac{\delta v}{\delta\rho}|HF>\rho\right),
\end{equation}
where  $v$ is the density-dependent part of the interaction.
We can deduce a generalized single-particle sum from that by
decomposing the second $\rho$ in the rearrangement term into
single-particle densities as $\rho=\sum_\alpha\rho_\alpha$. This
yields
\begin{equation}
 E=\frac{1}{2}\sum_{\alpha}w_{\alpha}(t_{\alpha}+\epsilon_{\alpha} + \tilde{\epsilon}_{\alpha})\;,
\end{equation}
where $\tilde{\epsilon}_{\alpha}$ is the single-particle rearrangement
energy arising from the above modification.  All of the binding
energies comprising the ion-ion interaction potential via
Eq.~(\ref{eq:vr}) can be expressed using the above form.  In terms of
these expressions the ion-ion interaction potential can be represented
as a sum over single-particle interaction potentials as
\begin{equation}
 V(R)=\sum_{\alpha}w_{\alpha} v_{\alpha}(R)\;,
\end{equation}
with
\begin{equation}
  v_{\alpha}(R)
  =
  \frac{1}{2}\left[(t_{\alpha}+\epsilon_{\alpha}+\tilde{\epsilon}_{\alpha})(R)-(t_{\alpha}^{0}+\epsilon_{\alpha}^{0}+\tilde{\epsilon}_{\alpha}^{0})\right],
\end{equation}
where the quantities with a superscript $(0)$ denote the asymptotic
values.  This decomposition is sensible if the single particle
energies are good representatives of the corresponding separation
energies according to Koopmans' theorem \cite{Koo33a} and if the
rearrangement corrections remain small. Such a situation is given in
the application example discussed below, namely nuclear mean field
calculations using Skyrme forces. In that case, the proton and neutron
interaction potentials are obtained simply by summing the
single-particle potentials $v_\alpha(R)$ over all protons, or neutrons
respectively. For electronic energy-density functionals one has to
employ a self-interaction correction to restore  Koopmans' theorem 
approximately \cite{Per81a,Mes08a}.
\begin{figure}[!htb]
\begin{center}
\includegraphics*[width=8.6cm]{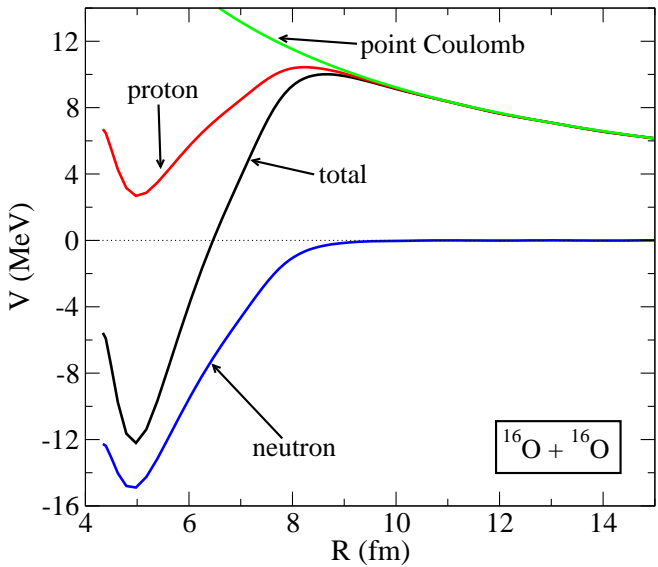}
\caption{\label{fig1} (color online) Interaction potential
for a head-on collision of $^{16}$O+$^{16}$O
at $E_{\mathrm{c.m.}}=12$~MeV. The black curve is the total ion-ion interaction potential. The red and blue curves show
the contribution to the total potential coming from protons and neutrons, respectively. Also shown is the point Coulomb
potential.}
\end{center}
\end{figure}

For the application of the above concepts we have used the DC-TDHF
approach~\cite{UO06b}.  In this approach TDHF time-evolution takes
place with no restrictions.  At certain times during the evolution the
instantaneous densities $\rho_{p/n}(\mathbf{r},t)$ are used to
perform a static Hartree-Fock minimization while holding the neutron
and proton densities constrained to be this instantaneous
TDHF densities~\cite{CR85}.  In essence, this introduces the concept
of an adiabatic reference state for a given TDHF state. The difference
between these two energies represents the internal energy.  The
adiabatic reference state is the one obtained via the density
constraint calculation, which is the Slater determinant with lowest
energy for the given density with vanishing current and approximates
the collective potential energy after the subtraction of the static
binding energies as shown in Eq.~(\ref{eq:vr}).  All of the dynamical
features included in TDHF are naturally included in the DC-TDHF
calculations.
\begin{figure}[!hbt]
\begin{center}
\includegraphics*[width=8.6cm]{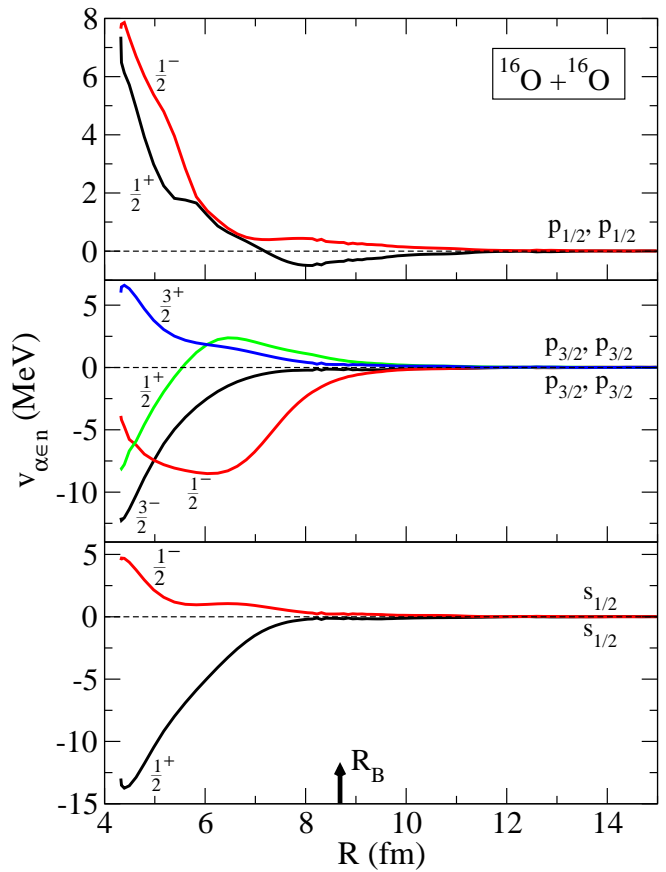}
\caption{\label{fig2} (color online) Contribution of neutron single-particles to the interaction potential
for a head-on collision of $^{16}$O+$^{16}$O
at $E_{\mathrm{c.m.}}=12$~MeV. The sum of all these potentials give the neutron curve in Fig.~\ref{fig1}.
The arrow marked $R_B$ indicates the barrier top of the total potential shown in  Fig.~\ref{fig1}.}
\end{center}
\end{figure}
\begin{figure}[!hbt]
\begin{center}
\includegraphics*[width=8.6cm]{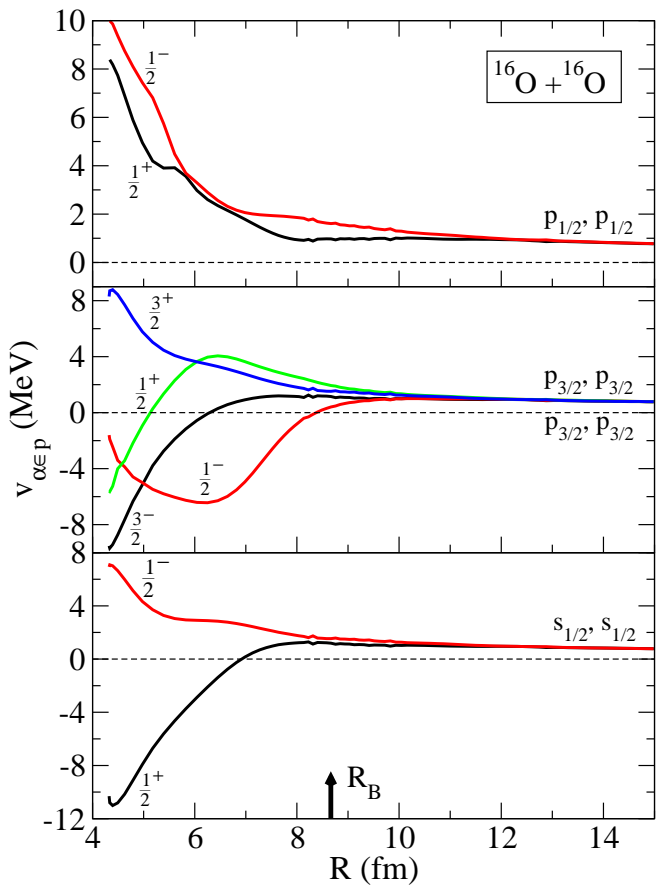}
\caption{\label{fig3} (color online) 
Same as Figure~\ref{fig2}, but for protons.}
\end{center}
\end{figure}

In practice, we have used the Skyrme SLy4 effective nuclear
interaction~\cite{CB98} for our calculations.  The rearrangement terms
described above can be written as;
\begin{equation}
  \tilde{\epsilon}_{\alpha} =  \tilde{\epsilon}_{\alpha}^{(3)}+ \tilde{\epsilon}_{\alpha}^{(C)}+\tilde{\epsilon}_{\alpha}^{(DC)}\;,
\end{equation}
with
\begin{eqnarray*}
 \tilde{\epsilon}_{\alpha}^{(3)}&=&-\frac{p}{12}t_3\int d^3r\;\rho_{\alpha}\rho^{p-1}\left[(1+\frac{x_3}{2})\rho^2\right. \\
 &\ &\qquad
 -\left.(\frac{1}{2}+x_3)(\rho_n^2+\rho_p^2)\right] 
\\
 \tilde{\epsilon}_{\alpha}^{(C)} 
 &=&
 -\frac{1}{2}\left(\frac{3}{\pi}\right)^{\frac{1}{3}}e^2
  \int d^3r\;\rho_{\alpha}\rho_p^{\frac{1}{3}} 
  \;,\;\alpha\in\{\mathrm{protons}\}
\\
  \tilde{\epsilon}_{\alpha}^{(DC)}&=& -\frac{1}{2}\int d^3r\;\rho_{\alpha}\lambda_{n,p}(\mathbf{r})\;,
\end{eqnarray*}
where we have defined
$\rho_{\alpha}=\psi_{\alpha}^{*}(\mathbf{r})\psi_{\alpha}(\mathbf{r})$,
and $\lambda(\mathbf{r})$ is the coordinate dependent Lagrange
multiplier for the density constraint~\cite{CR85}. The subscripts $n,p$ in the
rearrangement term for the density constraint indicates the choice
corresponding to the isospin content of the index $\alpha$.  The
summation over $\alpha$ exactly reproduces the total rearrangement
energies obtained by direct integration.

The first collision studied is $^{16}$O+$^{16}$O at $E_{c.m.}=12$~MeV.
In Fig.~\ref{fig1} we show the total interaction potential (black
curve) as well as the contributions coming from neutrons and protons
to this total potential.  
The total interaction potential is numerically identical to the one
obtained in standard DC-TDHF calculations using the  integral of the energy density to obtain the
terms in Eq.~(\ref{eq:vr}).
As a reference we also show the point Coulomb interaction.  As can be
seen from the figure the outer part of the potential barrier is
primarily determined by the interaction potential between the protons
of the two nuclei while the neutron potential is essentially zero. In
other words the neutron-proton interaction does not influence the
outer barrier region.  While inside the barrier neutrons provide all
of the attraction and the proton potential remains positive for all
$R$ values.
\begin{figure}[!htb]
\begin{center}
\includegraphics*[width=8.4cm]{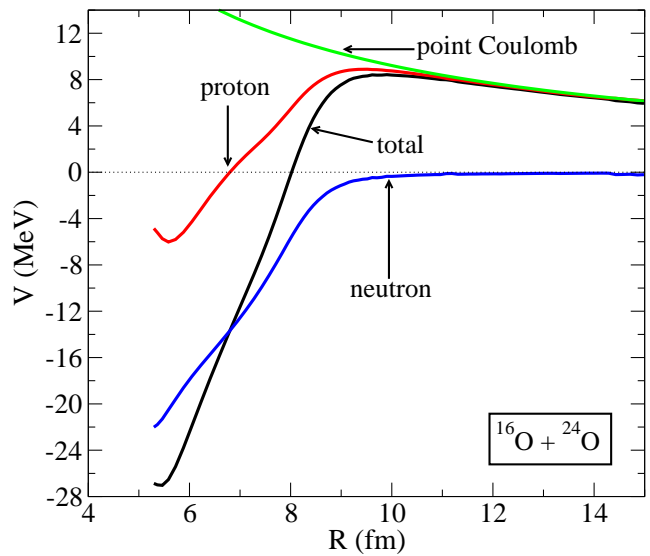}
\caption{\label{fig4} (color online) Interaction potential
for a head-on collision of the $^{16}$O+$^{24}$O
at $E_{\mathrm{c.m.}}=12$~MeV.}
\end{center}
\end{figure}

In Fig.~\ref{fig2} we show the individual neutron single-particle
potentials for the $^{16}$O+$^{16}$O system.  In spite of using
three-dimensional Cartesian coordinates, we are able to calculate some
quantum numbers when the solution possesses a good symmetry. In this
case spherical initial $^{16}$O nuclei allow the use of spherical
quantum labels at the asymptotic position of the two nuclei. On the
other hand, when the nuclei overlap the system becomes approximately
axially symmetric thus allowing the computation of parity and the
z-component of total angular momentum.  The reassignment of parities
and the behavior of the states as they evolve towards smaller $R$
values is noteworthy. For smallest $R$ values the filling of the
levels resembles the one for a single-centered $32$ particle system.
The trend seen in the actual single-particle energies is relatively
easy to understand; the two oxygen nuclei are initially far apart and
are brought together. While they are far apart and isolated the nuclei
have identical energy levels. However, as the separation between the
two nuclei becomes smaller the single-particle states begin to
overlap. The Pauli principle dictates that no two nucleons in an
interacting system may have the same quantum state. Therefore, each
(doubled) energy level of the isolated nuclei splits into two
orbitals, one lower in energy than the original level and one higher.
This can be most easily seen for the two initial $1s_{1/2}$ states.
However, in addition to energy splitting the parity
of one of these states also changes from positive to negative since no
more than two neutrons can be in the $1s_{1/2}$ state (each state is
originally occupied by two neutrons, $w_{\alpha}=2$).  In the language
of molecular physics the states that attain a lower energy than their
asymptotic value are referred to as the bonding states, whereas the
states that evolve to a higher energy are the anti-bonding
states \cite{Wei78aB}. We observe that not all neutron states are
bonding states but to the contrary about half are actually
anti-bonding in character for small $R$ values.  One of the $p_{3/2}$
states (green curve) make a transition from anti-bonding to bonding
for smaller $R$ values.  Another manifestation of the bonding and
anti-bonding states is their spatial localization. Bonding states
are localized in the interior part of the combined system as opposed
to anti-bonding states that are more spread out, as can been seen from
the single-particle moments.
Figure \ref{fig3} shows the contribution of proton single-particle
states to the total ion-ion potential.  The trend of the proton states
are essentially the same as the neutron states with the exception of
the rise from the zero potential line (dotted lines) due to the
presence of the Coulomb interaction.  The sum of all these proton
single-particle potentials reproduces the point Coulomb potential for
the two incoming nuclei as can been observed in Fig.~\ref{fig1}.  We
have also performed calculations for head-on collisions of
$^{16}$O+$^{24}$O at $E_{\mathrm{c.m.}}=12$~MeV, and for
$^{40}$Ca+$^{40}$Ca  at $E_{\mathrm{c.m.}}=55$~MeV.  In
Fig.~\ref{fig4} we show the neutron and proton potentials for the
$^{16}$O+$^{24}$O system. 
\begin{figure}[!htb]
\begin{center}
\includegraphics*[width=8.4cm]{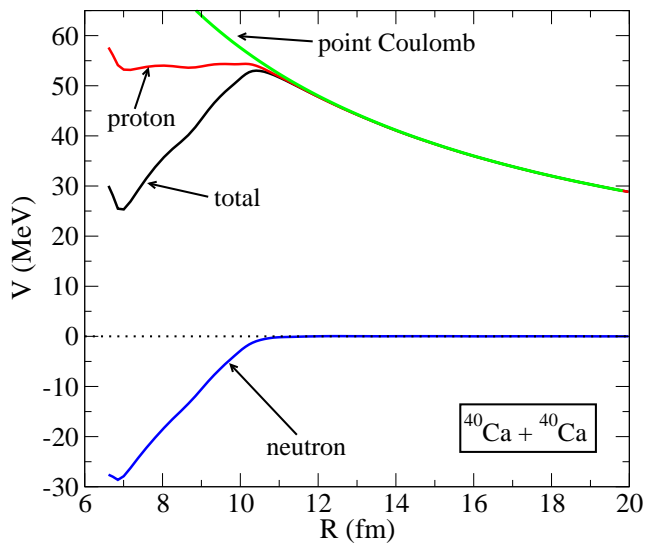}
\caption{\label{fig5} (color online) 
Same as Figure~\ref{fig4} but for
a head-on collision of the $^{40}$Ca+$^{40}$Ca
at $E_{\mathrm{c.m.}}=55$~MeV. }
\end{center}
\end{figure}
We note that in comparison to the
$^{16}$O+$^{16}$O system the potential minimum is considerably lower
due to the presence of eight extra neutrons. The surplus bending from
the extra neutrons also affect the proton potential, bringing it down
to negative values for small $R$.  We also note that the neutron
potential starts its dip to negative values at larger $R$ values than
the $^{16}$O+$^{16}$O case thus bringing the total barrier maximum to
a lower value.  Figure~\ref{fig5} shows the same quantities for the
$^{40}$Ca+$^{40}$Ca system.  The behavior of the proton potential is
significantly different in this case. Inside the barrier the proton
potential is essentially constant and it only rises as the minimum
value of $R$ is reached.  This is probably due to the stronger Coulomb
potential which counter-weights the nuclear binding.  While the
neutron and total potentials appear intuitively as expected, the
proton potential behaves in an unexpected manner.  This is further
evidence that the total potential may be masking some interesting
features of its building blocks.

We have introduced a general approach for the calculation of
single-particle interaction potentials as the building blocks of the
total interaction potential for two fragments of finite fermion
systems. The formulation is general provided the Koopmans' theorem is
applicable to the underlying energy functional used for the many-body
calculations.  The formalism is applied to calculate ion-ion
potentials for nuclear reactions using TDHF for the time-evolution of
the nuclear collision together with the density constraint formalism
to find the corresponding adiabatic reference state.  We show the
contribution of single-particle potentials to the total potential for
the $^{16}$O+$^{16}$O case. We identify repulsive and attractive
contributions as bonding and anti-bonding states.  Perhaps the more
obvious aspect is the identification of neutron and proton
contributions to the total potential where we see clearly the subtle
interplay between Coulomb repulsion and nuclear attraction for the
protons while neutrons are always the dominant contributors to binding.

This work has been supported by the U.S. Department of Energy under grant No.
DE-FG02-96ER40963 with Vanderbilt University, and by the German BMBF
under contract Nos. 06FY9086 and 06ER142D.

\end{document}